\documentclass[sn-nature]{sn-jnl}

\usepackage{graphicx}%
\usepackage{multirow}%
\usepackage{amsmath,amssymb,amsfonts}%
\usepackage{mathrsfs}%
\usepackage{xcolor}%
\usepackage{bm}
\usepackage{ragged2e}
\usepackage{mathptmx}  
\usepackage{siunitx}    
\renewcommand{\eqref}[1]{\hyperref[{#1}]{Eq.~\textup{(\ref*{#1}})}}
\newcommand{\FigRef}[1]{\hyperref[{#1}]{\textup{Fig.~\ref*{#1}}}}
\newcommand{\secref}[1]{\hyperref[{#1}]{\textup{Sec.~\ref*{#1}}}}
\newcommand{\tabref}[1]{\hyperref[{#1}]{\textup{Table~\ref*{#1}}}}
\makeatletter
\newcommand\figref[1]{%
    \@ifnextchar\bgroup{\figref@double{#1}}{\figref@single{#1}}%
}
\newcommand\figref@single[1]{\hyperref[{#1}]{\textup{Fig.~\ref*{#1}}}~}
\newcommand\figref@double[2]{\hyperref[{#1}]{\textup{Fig.~\ref*{#1}}\textcolor{blue}{\textbf{#2}}}}
\makeatother

\begin{document}
\title{\centering\textbf{Dispersive gains enhance wireless power transfer \\ with asymmetric resonance}}

\author[1,2]{\fnm{Xianglin} \sur{Hao}}\email{xianglhao2-c@my.cityu.edu.hk}
\equalcont{These authors contributed equally to this work.}

\author[2,3]{\fnm{Ke} \sur{Yin}}\email{yinke@scu.edu.cn}
\equalcont{These authors contributed equally to this work.}

\author[2]{\fnm{Shiqing} \sur{Cai}}\email{caisq1024@stu.xjtu.edu.cn}
\equalcont{These authors contributed equally to this work.}

\author[2]{\fnm{Jianlong} \sur{Zou}}\email{superzou@xjtu.edu.cn}

\author[2]{\fnm{Ruibin} \sur{Wang}}\email{3121104276@stu.xjtu.edu.cn}

\author[2]{\fnm{Xikui} \sur{Ma}}\email{maxikui@xjtu.edu.cn}

\author*[1]{\fnm{Chi K.} \sur{Tse}}\email{chitse@cityu.edu.hk}

\author*[2]{\fnm{Tianyu} \sur{Dong}}\email{tydong@mail.xjtu.edu.cn}

\affil*[1]{\orgdiv{Department of Electrical Engineering}, \orgname{City University of Hong Kong}, \orgaddress{\city{Hong Kong},  \country{China}}}

\affil*[2]{\orgdiv{School of Electrical Engineering}, \orgname{Xi'an Jiaotong University}, \orgaddress{\city{Xi'an}, \postcode{710049}, \state{Shaanxi}, \country{China}}}

\affil[3]{\orgdiv{College of Electronics and Information Engineering}, \orgname{Sichuan University}, \orgaddress{\city{Chengdu}, \postcode{610101}, \state{Sichuan}, \country{China}}}

\abstract{Parity-time symmetry is a fundamental concept in non-Hermitian physics that has recently gained attention for its potential in engineering advanced electronic systems and achieving robust wireless power transfer even in the presence of disturbances, through the incorporation of nonlinearity. However, the current parity-time-symmetric scheme falls short of achieving the theoretical maximum efficiency of wireless power transfer and faces challenges when applied to non-resistive loads. In this study, we propose a theoretical framework and provide experimental evidence demonstrating that asymmetric resonance, based on dispersive gain, can greatly enhance the efficiency of wireless power transfer beyond the limits of symmetric approaches. By leveraging the gain spectrum interleaving resulting from dispersion, we observe a mode switching phenomenon in asymmetric systems similar to the symmetry-breaking effect. This phenomenon reshapes the distribution of resonance energy and enables more efficient wireless power transfer compared to conventional methods. Our findings open up new possibilities for harnessing dispersion effects in various domains such as electronics, microwaves, and optics. This work represents a significant step towards exploiting dispersion as a means to optimize wireless power transfer and lays the foundation for future advancements in these fields.}



\maketitle

\section{Introduction} \label{sec:intro} 
The concept of wireless power transfer (WPT) can be traced back to the groundbreaking work of Nicola Tesla in the late 19th century \cite{song2021wireless}. Over the past two decades, numerous WPT platforms have been proposed, including microwaves \cite{hajimiri2020dynamic,ayling2024wireless}, acoustics \cite{tseng2017phased,bakhtiari2018acoustic}, and non-radiative methods \cite{kurs2007wireless,assawaworrarit2017robust,xie2021magnetic,zanganeh2021nonradiating}. Among these platforms, near-field magnetic resonant coupling wireless power transfer has gained significant attention due to its high efficiency, power capacity, and flexibility \cite{kurs2007wireless}. In recent years, the application of modern non-Hermitian physics principles, such as parity-time (PT) symmetry, has elevated the robustness of WPT systems to new heights \cite{assawaworrarit2017robust,sakhdari2020robust}. PT symmetry has provided solutions to the challenges associated with large-scale WPT, addressing concerns such as cost, efficiency, and resistance to disturbances. However, the reliance on symmetry also imposes stringent requirements on system parameter design \cite{hao2023frequency}, thereby limiting the further advancement of WPT technologies. Furthermore, the strict adherence to PT symmetry may be impractical for commonly used batteries or rectifier loads \cite{wang2024enhanced}, necessitating the consideration of asymmetric scenarios in engineering practice. 

In practical applications, asymmetric designs offer several advantages, such as the elimination of the need for precise tuning, reduced production costs, and improved efficiency for fixed-frequency WPT schemes \cite{zhang2013design}. However, in the context of non-Hermitian self-oscillation WPT systems, the multi-mode characteristics and steady-state mechanism pose challenges for achieving high transfer efficiency and power stability in asymmetric structures \cite{hao2023frequency}, making it difficult to promote their practical applications. In this study, we present an enhanced asymmetric resonant circuit that incorporates {\em dispersive nonlinear saturable} gain elements. We demonstrate that by combining non-Hermitian physics, nonlinearity, and dispersion, it is possible to achieve steady-state mode selection. This allows us to operate at frequencies that offer optimal efficiency, surpassing the limitations of PT-symmetric systems. The elimination of the requirement for symmetry in the natural resonant frequency enables the connection of various loads and stretchable electronic resonators \cite{choi2022transient,kim2024strain} to asymmetric WPT systems, particularly beneficial for scenarios where frequency changes are likely to occur. 

Our contribution extends beyond proposing a practical design path; it delves into the unexplored realm where dispersion phenomena and non-Hermitian physics intertwine, offering significant multidisciplinary influence and inspiration. Dispersion, a fundamental phenomenon in wave systems, has garnered extensive research in various fields of physics. For instance, in optics, chromatic dispersion refers to the frequency dependence of the refractive index of a material \cite{born2013principles,mcclung2020will,chen2020flat}. However, in the context of non-Hermitian (PT-symmetric) optical systems \cite{phang2014impact,phang2015parity,nguyen2016recovering,shramkova2018dispersive}, dispersion is typically perceived as an undesirable factor due to its tendency to limit the frequency band \cite{phang2014impact}.Similar to wave systems, dispersion is a fundamental phenomenon that also manifests in circuit systems, where circuit parameters, including resistance, exhibit frequency dependence in high-speed design \cite{bogatin2010signal}. Furthermore, the study of non-linear saturation, a well-established phenomenon, has extensively focused on non-Hermitian circuits \cite{hu2010nonlinear,assawaworrarit2017robust,zhu2022anomalous,xia2021nonlinear}. However, the combined effects of dispersion and nonlinear saturation have remained largely unexplored. Building upon pioneering works in wave and circuit systems, we have developed a frequency-dependent negative resistor with nonlinear saturation using a generalized impedance converter. By investigating the implications of non-Hermitian systems with dispersive nonlinear gain in asymmetric WPT systems, we have made significant advancements in this field. Remarkably, we have discovered that the introduction of dispersive gain can greatly optimize the energy distribution within the system, resulting in concentrated energy primarily on the load side. This optimization effectively reduces the losses in the source resonator, leading to higher power transfer efficiency compared to PT-symmetric cases. Our findings highlight the potential of dispersive nonlinear gain in enhancing the performance of asymmetric WPT systems, surpassing the limitations of traditional symmetric approaches.

\section{Results} \label{sec:results}
\subsection{Concepts of enhanced asymmetric resonance} \label{sec:concepts}
\begin{figure}[!ht]
    \centering
    \includegraphics[width=\linewidth]{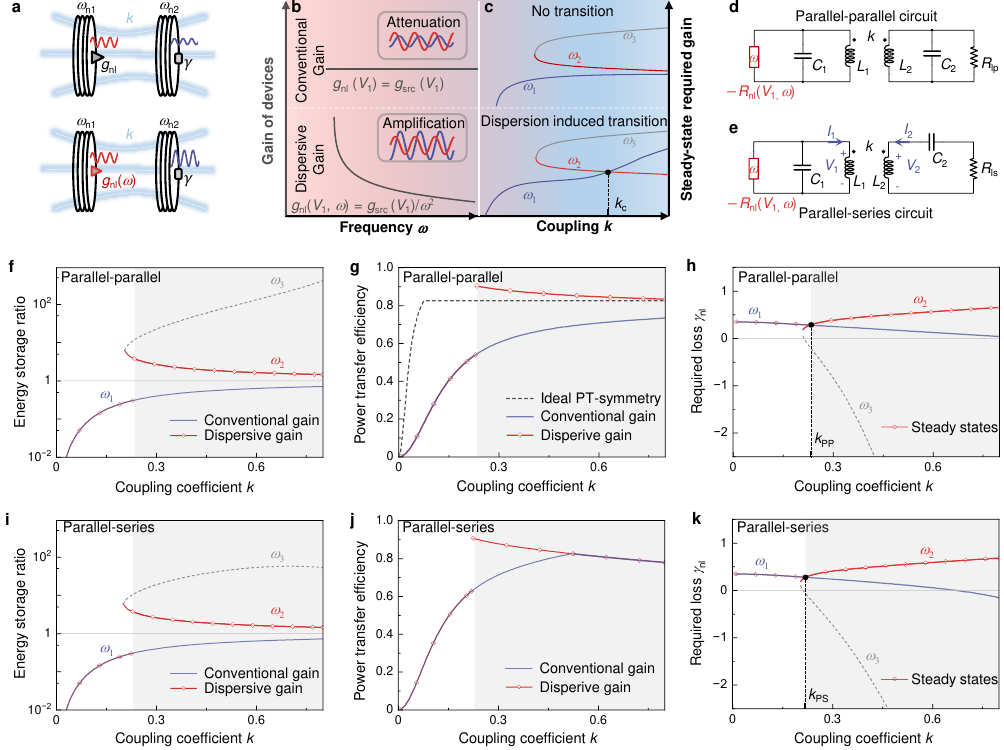}
    \caption{\textbf{Comparison between conventional-gain-based and dispersive-gain-based wireless power transfer schemes as the coupling coefficient $k$ varies.} \textbf{a},~Conventional saturation-gain and dispersion-gain schemes. \textbf{b},\textbf{c},~Comparison of \textbf{b} devices' gain and \textbf{c} steady-state required gain spectrum. Circuit diagrams of the coupled resonator system using \textbf{d} parallel-parallel and \textbf{e} parallel-series topologies. \textbf{f},~Calculated energy storage ratio $|a_2/a_1|^2$, \textbf{g}, power transfer efficiency and \textbf{h},~required steady-state losses $\gamma_\text{nl}$ as functions of the coupling parameter $k$ of the parallel-parallel circuit \textbf{d}. \textbf{i}\,--\,\textbf{k} correspond to \textbf{f}\,--\,\textbf{h} but for the parallel-series topology \textbf{e}. For all figures, the asymmetry parameter is $\chi = 4/3$. Shaded regions in \textbf{f}\,--\,\textbf{h} and \textbf{i}\,--\,\textbf{k} denote the strongly coupling region, which are determined by the intersection points of the desired steady-state loss \eqref{eq:disGY} spectrum in \textbf{h} and \textbf{k}, respectively. }
    \label{fig:fig01}
\end{figure}

To elucidate the enhancement of wireless power transfer in an asymmetrical scheme through the use of dispersive gains, we consider an inductively coupled two-resonator system with a coupling coefficient $k$, as illustrated in \figref{fig:fig01}{a}. The resonators in the system, namely the source and receiver resonators, possess natural resonance frequencies of $\omega_\text{n1}$ and $\omega_\text{n2}= \omega_\text{n1} / \sqrt{\chi}$, respectively. Here, $\chi$ represents the ratio of the natural resonant frequencies and incorporates the asymmetry. The resonators are subject to a nonlinear dispersive gain $g(\omega) = g_\text{nl}(V_1,\omega)$ and a loss coefficient $\gamma$.
In addition to the asymmetry in the natural resonant frequencies, the steady-state mechanism and performance of the system are influenced by the microstructure of the circuit resonator, specifically whether it is in a parallel or series configuration. \figref{fig:fig01}{d} and \figref{fig:fig01}{e} illustrate the parallel-parallel and parallel-series microstructures, respectively.

We now proceed with the analysis for the parallel-series (PS) structure, depicted in \figref{fig:fig01}{d}, as an example. Similar approaches are applicable for the analysis of other structures, which are detailed in Section 1 in the Supplementary Information. Having defined the state vector $\bm{\Psi} = (V_1, V_2, I_1, I_2)^\text{T}$ and $\tau = \omega_0 t$ with $\omega_0 = \omega_\text{n2}$, the system dynamics can be described by $\text{d}\bm{\Psi}/\text{d}\tau = \mathscr{L}_\text{PS} \bm{\Psi}$ with the Liouvillian being
\begin{equation} \label{eq:dynL}
    \mathscr{L}_\text{PS} = 
    \begin{pmatrix}
        \chi g_\text{nl} (V_1, \omega) & 0 & -\chi Z_\text{r}  & 0 \\
        \cfrac{k}{1-k^2} \gamma & -\cfrac{1}{1-k^2} \gamma & 0 & -Z_\text{r} \\
        \cfrac{1}{1-k^2} \cfrac{1}{Z_\text{r}} & -\cfrac{k}{1-k^2} \cfrac{1}{Z_\text{r}} & 0 & 0 \\
        -\cfrac{k}{1-k^2} \cfrac{1}{Z_\text{r}} & \cfrac{1}{1-k^2} \cfrac{1}{Z_\text{r}} & 0 & 0 \\
    \end{pmatrix},
\end{equation} 
where $g_\text{nl}(V_1, \omega)$ is the nonlinear gain of the dispersive source devices, which is related to the operating frequency $\omega$. Here, $V_n$ and $I_n$ respectively denote the voltage and current amplitudes over and across the source (or the inductor on the load side). Subscripts $n = 1$ and $n = 2$ correspond to the source and receiver, respectively. In addition, $\omega$ is the frequency normalized by the natural resonant frequency $\omega_0 = \omega_\text{n2}$ of the receiver resonator and $Z_\text{r}=\sqrt{L_2/C_2}$ denotes the wave impedance of the receiver resonator with $L_2$ and $C_2$ being the inductance and capacitance, respectively. The loss parameter is defined by $\gamma=R_\text{ls}/Z_r$ for the PS circuit. 

For a given nonlinear gain $g_\text{nl}(V_1, \omega)$, the operating frequency $\omega$ in the steady state must satisfy the characteristic equation $\det(\omega I - \text{i} \mathscr{L}_\text{PS}) = 0$, i.e.,
\begin{equation} \label{eq:chact}
    \omega ^4 + \frac{1 +\chi - \chi \gamma g_\text{nl} (V_1, \omega)}{k^2-1} \omega^2 +\frac{\chi }{1-k^2}
    + \text{i} \left\{ \left[ \chi g_\text{nl} (V_1, \omega) +\frac{\gamma }{k^2-1} \right] \omega^3 + \chi \frac{g_\text{nl}(V_1, \omega)-\gamma}{k^2-1} \omega \right\} = 0.
\end{equation}
Alternatively, the particular nonlinear gain $g_\text{nl,ss}$ required to maintain stability with a real-valued operating frequency $\omega$ can be obtained by setting the imaginary part of \eqref{eq:chact} to zero, i.e.,
\begin{equation} \label{eq:gain}
    g_\text{nl,ss}  = \frac{\gamma}{\chi} \frac{1}{k^2-1} \left[ \frac{(k^2-1)\chi + 1}{(k^2-1)\omega^2+1} - 1 \right].
\end{equation}
Thus, the required nonlinear gain $g_\text{nl,ss}$ depends on the coupling coefficient $k$ and the operating frequency $\omega$ for a given loss parameter $\gamma$ and an asymmetry parameter $\chi$. Since non-Hermitian systems can be self-oscillating \cite{ra2018site,liu2019pulsed,li2023synergetic}, by putting the real part of \eqref{eq:chact} to zero and considering \eqref{eq:gain}, we can obtain the characteristic equation, i.e., $\omega^6 + c_1 \omega^4 + c_2 \omega^2 + c_3 = 0$, where $c_1 = [\gamma^2 + (k^2-1)(\chi+1)]$, $c_2 = [1 - (\gamma ^2+k^2-2)\chi]/(k^2-1)^2$, and $c_3 = -\chi/(k^2-1)^2$. Consequently, the eigenfrequencies (potential supported modes) can be obtained as
\begin{subequations} \label{eq:fre}
\begin{align}
    \omega_1 &= \sqrt{-c_1/3 + t_1 + t_2}, \label{eq:fre1} \\
   \omega_{2,3} &= \sqrt{-c_1/3 - (t_1+t_2)/2 \pm \text{i} \sqrt{3} (t_1-t_2)/2 }, \label{eq:fre2} 
\end{align}
\end{subequations}
where $t_1=\sqrt[3]{p+\sqrt{p^2+q^3}}$ and $t_2=\sqrt[3]{p -\sqrt{p^2+q^3}}$ with $p=-c_1^3/27 +c_1 c_2 /6 -c_3/2$ and $q= -c_1^2/9 +c_2/3$. Similar to the perturbative PT symmetry scheme \cite{yin2022wireless}, the eigenfrequencies given in \eqref{eq:fre} are functions of the loss parameter $\gamma$, the coupling coefficient $k$, and the asymmetry parameter $\chi$. For the parallel-parallel (PP) circuit topology, the system can degenerate into the PT symmetric case when $\chi = 1$ (see Section 2.1 of Supplementary Information). Moreover, the energy distribution in the resonators for different supportable modes can be identified using the coupled mode theory, which is based on the energy stored $|a_n|^2 = C_n V_{C_n}^2/2$ in the capacitor, where $V_{C_n}$ denotes the voltage across the capacitor $C_n$. A higher ratio of $|a_2/a_1|^2$ indicates more energy stored in the load resonator, as shown in \figref{fig:fig01}{f} and \figref{fig:fig01}{i} for the PP and PS structure, respectively.

According to \eqref{eq:fre}, although multiple modes are supported, they may not exist simultaneously. Previous works \cite{yin2022wireless,assawaworrarit2017robust,hao2023frequency} have demonstrated that the steady-state selection mechanism of the non-Hermitian multi-mode system follows the principle of minimum gain \cite{assawaworrarit2017robust}. Specifically, for a coupled resonant circuit with a single nonlinear saturated gain whose initial value is zero, the mode requiring the lowest gain will be the final steady-state mode. Now, substituting the eigenfrequencies given in \eqref{eq:fre} into the required nonlinear gains calculated from \eqref{eq:gain}, the steady-state mode of the system can be determined by finding the minimum gains of \eqref{eq:gain} from the three modes \eqref{eq:fre}. \figref{fig:fig01}{b} illustrates the conventional non-dispersive and frequency-dependent saturation gains of the devices, accompanied by the evolution of the required steady-state gain as the coupling coefficient varies, as shown in \figref{fig:fig01}{c}. The conventional gain system always operates in the $\omega_1$ mode, since the corresponding gain is always the minimum. However, as shown in \figref{fig:fig01}{f}, for a parallel-parallel circuit adopting the conventional non-Hermitian scheme, the energy storage ratio for $\omega_1$ is the lowest among the three modes and is always smaller than one, which is the energy ratio for the PT-symmetry scheme, indicating that the asymmetric design fails to improve efficiency as in the fixed frequency design \cite{zhang2013design} and the system efficiency drops to a value inferior to the PT-symmetric case. Is it possible to exploit the other two theoretical modes $\omega_{2,3}$ which have higher energy storage ratios compared to the PT symmetry case with possibly improved efficiency? Here, we provide an answer by exploiting the dispersion effect.

We first note that $g_\text{nl}(V_1, \omega)$ as a function of $V_1$ and $\omega$ for the actual gain device can take different forms, and a steady state can be achieved provided that the required steady-state value of $g_\text{nl,ss}$ calculated from \eqref{eq:gain} is within the range $g_\text{nl}(V_1, \omega)$. Thus, \eqref{eq:gain} determines the steady-state competition, while $g_\text{nl}(V_1, \omega)$ does not affect eigenfrequencies \eqref{eq:fre}, as restricted by \eqref{eq:gain}. Therefore, dispersive gain elements can be used to reshape the steady-state gain \eqref{eq:gain} and select the desired steady-state mode. In general, the system dynamics can be extremely complicated due to nonlinear saturation and dispersion. Here, we consider the case where nonlinearities and frequency dependencies are separable variables, i.e., $g_\text{nl}(V_1, \omega) = g_\text{src}(V_1) g_\text{d}(\omega)$. To get a concise and intuitive model, we assume an ideal case where the gain function of the source devices is $g_\text{nl}(V_1,\omega)= g_\text{src}(V_1)/\omega^2$, with $g_\text{src}(V_1)$ being the voltage-dependent gain of the device which exhibits saturation behavior. Now, the steady-state gain required for the device $g_\text{src}$ is reshaped as $g_\text{src,ss} = \omega^2 g_\text{nl,ss}$ according to \eqref{eq:gain}, i.e.,
\begin{equation}\label{eq:disG}
    g_\text{src,ss} = \frac{\gamma}{\chi} \frac{1}{k^2-1} \left[ \frac{(k^2-1)\chi + 1 }{(k^2-1) + 1/\omega^2} - \omega^2 \right].
\end{equation} 
As shown in \figref{fig:fig01}{c}, the dispersive gain curves have an intersection point $k = k_c$. The minimum gain mode is $\omega_1$ when $k < k_c$ and becomes $\omega_2$ when $k > k_c$, demonstrating a dispersion-induced transition for the proposed system. Moreover, this steady-state switching can re-distribute the energy, allowing the energy storage ratio of the system to be greater than one in the strongly coupling region, as shown in \figref{fig:fig01}{f} and \figref{fig:fig01}{i} for the PP and PS systems, respectively. The increment of the energy storage ratio to increase and hence improve the power transfer efficiency for symmetrical topologies such as PP circuits, as shown in \figref{fig:fig01}{g}. Using a dispersive gain, asymmetric resonance  ($\omega_\text{n1} \neq \omega_\text{n2}$) has a higher power transfer efficiency than PT-symmetry resonance in the strong coupling region, as shown in \figref{fig:fig01}{g}. However, for PS circuits, this may not be true since the loss rate of the resonator depends on the microstructures. Nonetheless, the dispersion-induced steady-state selection can also improve the efficiency of asymmetric WPT at specific parameters, as shown in \figref{fig:fig01}{j}. In addition, similar to the transition caused by dispersion, the circuit topology difference between the source and receiver can also trigger a transition (see Supplementary Information, section 2.2). In practice, a more feasible implementation is to combine the dispersive gain and the nonlinear loss (see Supplementary Information, section 3.1) since constructing an excellent gain element \eqref{eq:gain} can be a big challenge, which can be modeled as $g_\text{nl}(V_1,\omega) = g_\text{c}/\omega^2 - \gamma_\text{nl}(V_1)$ where $g_\text{c}$ is a constant value and the nonlinear loss $\gamma_\text{nl}(V_1)$ changes adaptively to maintain the steady state. Similar to \eqref{eq:disG}, the required steady-state loss is
\begin{equation}\label{eq:disGY}
    \gamma_\text{nl,ss} = \frac{g_\text{c}}{\omega ^2} + \frac{\gamma}{\chi} \frac{\omega^2 - \chi}{(k^2-1)\omega^2 + 1}.
\end{equation}
Unlike the gain, the mode with the greatest loss $\gamma_s$ is the steady state. \figref{fig:fig01}{h} and \figref{fig:fig01}{k} display the required loss spectrum of the dispersive-gain system for the PP and PS structures, respectively, showing a dispersion-induced transition which has an even wider strongly coupled region than the ideal dispersive gain system since $k_\text{PS}<k_\text{PP}<k_\text{c}$ (see Section 2 of Supplementary Information).

\subsection{Experimental verification}\label{subsec:results}
\begin{figure}[!t]
    \centering
    \includegraphics[width=\linewidth]{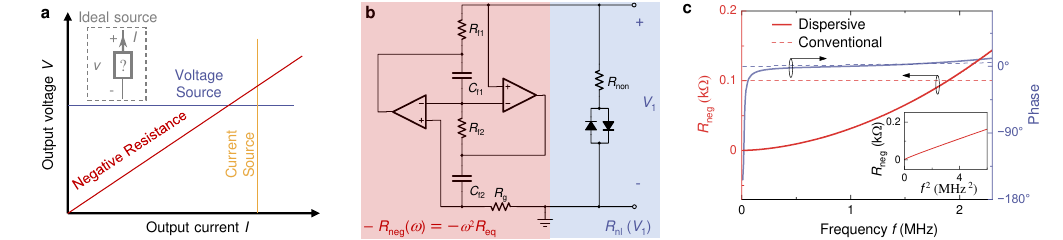}
    \caption{\textbf{a}, Comparison of ideal linear power supply models. \textbf{b}, Circuit diagram of the dispersive nonlinear gain element. \textbf{c}, Comparison of simulated negative resistance behavior between the dispersive element (solid curve) and conventional saturation gain element (dashed line).}
    \label{fig:fig02}
\end{figure}

As discussed above, a dispersive nonlinear gain element is necessary for efficiency improvement. In non-Hermitian schemes, negative resistors are commonly used as gain elements. \figref{fig:fig02}{a} compares the characteristics of the ideal negative resistors with conventional power supplies. Classic power supply models often lack the dispersion property and cannot support self-oscillation in the absence of nonlinearity. To incorporate the dispersion characteristics and the nonlinear behavior of negative resistors, we use a gain element consisting of a generalized impedance converter (GIC) \cite{mishonov2021q} which serves as a frequency-dependent negative resistance $-\omega^2 R_\text{eq}$ and a nonlinear resistor $R_\text{nl}(V_1)$ comprising of a resistor and anti-parallel diodes, where $R_\text{eq} = C_\text{f1} C_\text{f2} R_\text{f1} R_\text{f2} R_\text{g}$, as shown in \figref{fig:fig02}{b}. Consequently, the equivalent gain of this nonlinear element is approximately $g_\text{nl}(V_1,\omega) = g_\text{c}/\omega^2 - \gamma_\text{nl}(V_1)$ with $g_c =  \sqrt{L_2/C_2}/R_\text{eq}$ and $\gamma_\text{nl}(V_1) \propto R_\text{nl}^{-1}(V_1)$ being a nonlinear loss (see Fig. S6c in Supplementary Information), which can achieve the steady-state gain behavior corresponding to \eqref{eq:disGY}. Also, \figref{fig:fig02}{c} shows the simulated equivalent negative resistance as a function of frequency for the proposed frequency-dependent negative resistor and the traditional negative resistor. Here, the phase-frequency relation (solid blue curve) indicates that the output current and input voltage of the GIC element are basically in phase except for extremely low frequencies, providing a negative resistance characteristic. Moreover, the dispersive negative resistance designed is proportional to $\omega^2$ (see the inset of \figref{fig:fig02}{c}), while the conventional resistance is approximately independent of the frequency in the design frequency band (dashed lines). 

\begin{figure}[!t]
    \centering
    \includegraphics[width=\linewidth]{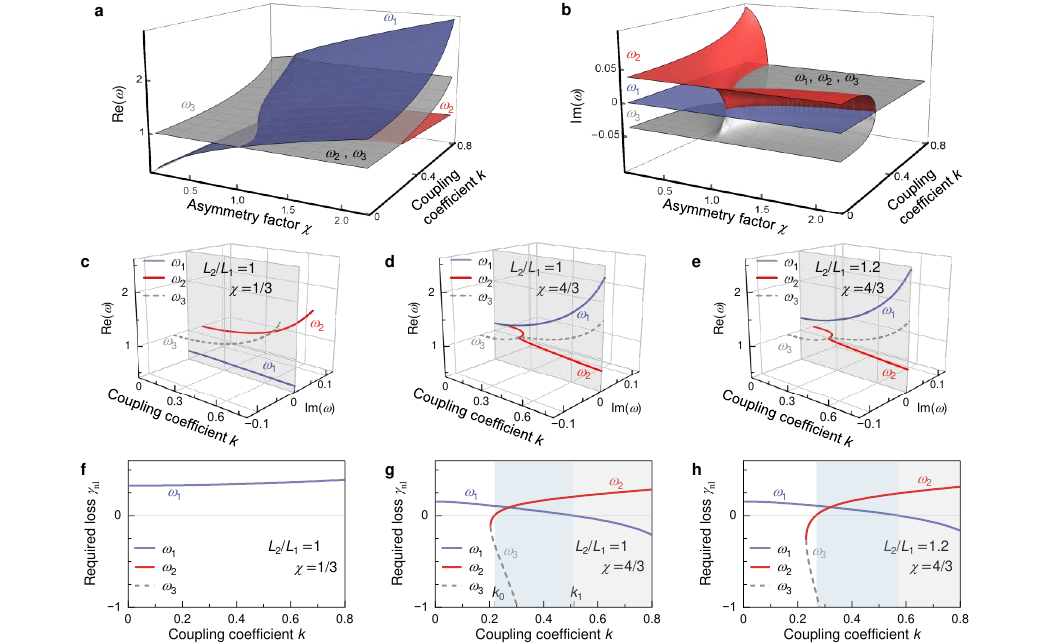}
    \caption{\textbf{Evolution of eigenfrequencies and steady-state gains of the dispersion-gain non-Hermitian system.} \textbf{a}, Real $\Re(\omega_n$) and \textbf{b}, Imaginary $\Im(\omega_n$), \textbf{b}) parts of the eigenfrequencies normalized by $\omega_0$ of the asymmetry-resonant system as a function of the detuning parameter (a.k.a. asymmetry factor) $\chi$ and the coupling coefficient $k$. \textbf{c\,--\,e}, Evolution of the computed eigenfrequencies with respect to $k$ for different parameters: \textbf{c}, $L_1/L_2 = 1$, $\chi = 1/3$; \textbf{d}, $L_1/L_2 = 1$, $\chi = 1/4$; and \textbf{e}, $L_1/L_2 = 1.2$, $\chi = 4/3$. \textbf{f\,--\,h}, Required losses correspond to \textbf{c--e}, respectively.}
    \label{fig:fig03}
\end{figure}

Before conducting the experiment, it is necessary to understand the impact of the detuning of the resonance frequency of resonators on the steady-state operation. \figref{fig:fig03} illustrates the complex eigenfrequencies of the PS circuit as functions of the coupling coefficient $k$ under different detuning ratios $\chi$. As shown in \figref{fig:fig03}{a} and \figref{fig:fig03}{b}, $\omega_1$ (the blue sheet) is always real, while $\omega_2$ and $\omega_3$ are real only when coupling is strong, as indicated by the flat regions in \figref{fig:fig03}{b} where the imaginary part of the eigenfrequency is zero. In particular, when $\chi \ll 1$, e.g., $\chi=1/3$, only one real eigenfrequency $\omega_1$ exists for all $k$ as shown in \figref{fig:fig03}{c}, and only $\omega_1$ mode is physically valid in the required steady-state gain spectrum (\figref{fig:fig03}{f}). Therefore, the dispersion-induced steady-state selection mechanism is more suitable for the case when $\chi>1$. For example, when $\chi=4/3$, $\omega_2$ and $\omega_3$ are real in the strongly coupled region and become complex conjugates in the weakly coupled region (\figref{fig:fig03}{d}). Also, a transition point exists, which is determined by the intersection of the gain curves instead of the imaginary part of the frequency domain (\figref{fig:fig03}{g}). Even when the inductances of the source and load resonators are slightly different, a similar mode frequency evolution and gain behavior can be retained, as shown in \figref{fig:fig03}{e} and \figref{fig:fig03}{h}, respectively.

\begin{figure}[!t]
    \centering
    \includegraphics[width=0.97\linewidth]{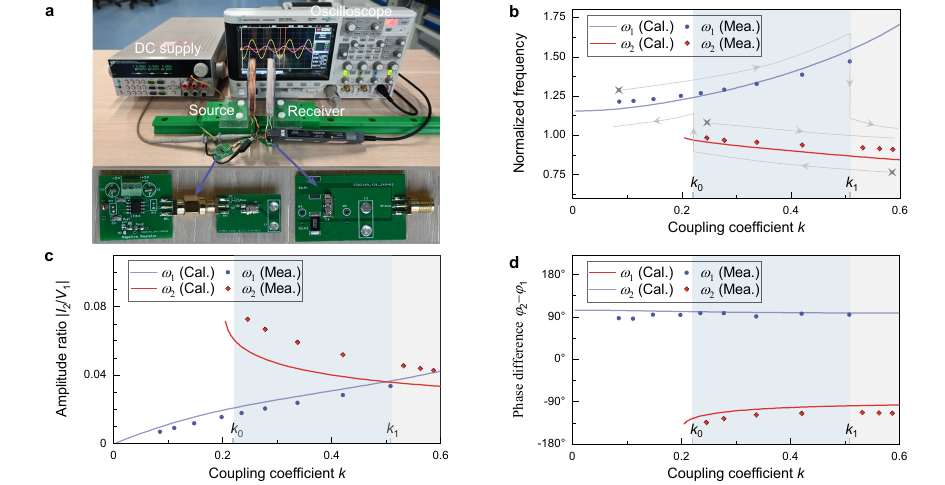}
    \caption{\textbf{Experimental verification of asymmetric-resonant wireless power transfer based on dispersive gain.} \textbf{a}, Photo of the experimental prototype. \textbf{b\,--\,d}, Comparison of theoretical and experimental results for the normalized frequency. \textbf{b}, amplitude ratio $|I_2/V_1|$, \textbf{c}, and \textbf{d}, phase difference $\varphi_2-\varphi_1$ behaviors as the coupling coefficient $k$ varies. The gray lines with arrows in \textbf{b} depicts the experimental situations where $k$ is increased or decreased at different starting points marked in gray stars. Note that $k_0$ and $k_1$ are the theoretical calculated values as the same as those in \FigRef{fig:fig03}\textbf{g}. The critical coupling coefficient $k_0$ of the strong coupling region depends on the intersection of the gain spectrum rather than the complex bifurcation point of the eigenfrequency.}
    \label{fig:fig04}
\end{figure}

\figref{fig:fig04}{a} shows a prototype of the proposed WPT system with a PS structure based on the detailed setup given in the previous section. The measured steady-state frequency shows good agreement as the calculated value for various coupling coefficients, as shown in \figref{fig:fig04}{b}. It is worth noting that the experimental observations confirm that the steady-state behavior of the system is related to the initial value of the nonlinear loss. When the system is powered on with very weak coupling when $k < k_0$, the system operates in steady-state mode $\omega_1$ and will not switch to mode $\omega_2$ until $k > k_1$ as $k$ gradually increases without being powered off. In contrast, when the system starts at a strong coupling when $k > k_0$, the system operates in steady-state mode $\omega_2$ and will not switch to mode $\omega_1$ until $k < k_0$ as $k$ gradually decreases. Here, the transition points $k = k_0$ and $k = k_1$ are respectively the intersections of $\gamma_\text{nl}(k) = 0$ for $\omega_2$ and $\omega_1$, as shown in \figref{fig:fig03}{g}. Interestingly, similar to the PT-symmetric system, a bistable region is present when $k_0< k < k_1$ \cite{wang2019dynamics,li2023synergetic,cui2022high}, which results in the asymmetric hysteretic mode switching \cite{wang2019dynamics} for different starting conditions (see Section 4 of Supplementary Information). \figref{fig:fig04}{c} and \figref{fig:fig04}{d} display the ratio of the load current $I_2$ to the source voltage $V_1$, and the relative phase between $I_2$ and $V_1$, respectively. The deviation between the measured and calculated values of the ratio $|I_2/V_1|$ is due to the high sensitivity of $|I_2/V_1|$ to frequency change, while the measured and calculated frequencies differ slightly. In fact, the higher measured amplitude ratio compared to the theoretical value of the mode $\omega_2$ is conducive to improving transmission efficiency and power (see (S9) in Supplementary Information). Unlike the frequency and amplitude ratios, the phase difference $\varphi_2-\varphi_1$ is relatively stable and insensitive to the coupling change. In addition, the significant difference in the phase difference between the two modes make it possible for detecting the operating status of the system.

\section{Discussion}\label{sec:discussion}
Our contribution is in enhancing asymmetric power transfer through adopting dispersion. Specifically, our work enables the computation and comparison of the steady-state properties of coupled resonator systems with various microstructures and gain types. We analyze the impact of the source resonator's quality factor on efficiency by concentrating the energy predominantly in the receiver resonator at the dispersion-induced transition-selected steady state (refer to Section 1 of Supplementary Information). Additionally, our qualitative analysis, based on the principle of minimum gain, offers designers significant flexibility. Designers can now envision a gain, even if it has not been experimentally observed, and evaluate whether incorporating this gain would enhance the system performance. The dispersion-induced transitions, which arise from the prevalent dispersion phenomenon in nature, introduce a novel mode transition mechanism. This mechanism holds immense interdisciplinary value and is expected to significantly advance optical systems \cite{nasari2022observation} and optomechanical systems \cite{xu2016topological}. In engineering applications, we anticipate further improvements in the overall efficiency and output power by employing gain elements constructed with switch-mode amplifiers \cite{assawaworrarit2020robust,zhou2018nonlinear,yang2024digital} and mitigating additional nonlinear losses. Fractional-order control \cite{jiang2019fractional} and digital feedback control \cite{yang2024digital} are potential approaches to develop a saturated dispersive gain with high efficiency.

\section{Conclusion} \label{sec:conclusions}
We have conducted an investigation into non-Hermitian electronic dimer systems featuring dispersive nonlinear gain. Our focus has been on understanding the steady-state mechanism and energy distribution within these systems. We have successfully demonstrated the attainment of enhanced efficiency in wireless power transfer at an asymmetric resonance. This breakthrough overcomes the efficiency limitations typically associated with non-Hermitian systems while maintaining a high level of robustness.
To validate our findings, we have carried out simulations and performed experimental measurements on a prototype, which provide substantial supporting evidence for our theoretical predictions. These results underscore the practical viability of our approach.
Significantly, we have introduced the concept of dispersive gain into the realm of non-Hermitian physics. Based on this dispersive gain, we have developed a mode selection method driven by dispersion, enabling the design of steady-state operation in multi-mode systems. This pioneering contribution expands the frontiers of non-Hermitian physics, opening new avenues for exploration and innovation in this field.

\section*{Method}\label{sec:methods}
\textbf{Measurement set-up.} 
The wireless power transfer prototype circuit is composed of two distinct units of the source and receiver resonators, as shown in \figref{fig:fig04}{a}. The two coils of source and receiver resonators are composed of four turns of Litz wire wrapped around a 10-cm diameter ring, resulting in self-inductances of $L_1 = 3.31~\si{\micro\henry}$ ($\phi0.2~\si{\milli\meter} \times 50$ strands where $\phi$ denotes the diameter of the wire) and $L_2 = 3.35~\si{\micro\henry}$ ($\phi 0.05~\si{\milli\meter} \times 660$ strands), respectively. To maintain alignment as the transfer distance varies, a nylon plastic rail is used to ensure the coaxial alignment of the two coils on the slider. The source coil is connected in parallel with a capacitor $C_1 = 3.3~\si{\nano\farad}$, resulting in a resonance frequency of $\omega_\text{n1} = 1.53~\si{\mega\hertz}$, and a measured quality factor of $Q_\text{s}=72$. The receiver resonator consists of a coil in series with a capacitor ($C_2 = 4.4~\si{\nano\farad}$) and a load resistor ($R_2 = 2~\si{\ohm}$), whose resonance frequency measured is $\omega_\text{n2} = 1.31~\si{\mega\hertz}$ and the quality factor is $Q_\text{r}=268$. The nonlinear dispersive gain element consists of a high-speed operational amplifier LT1813 with $R_\text{f1} = R_\text{f2} = 51~\si{\ohm}$ and $C_\text{f1} = C_\text{f2} = 2.2~\si{\nano\farad}$. The gain rate $g_c =  \sqrt{L_2/C_2}/R_\text{eq}$ can be finely reduced to an appropriate value with an adjustable feedback resistor $R_\text{g}$ ($R_\text{g} = 80~\si{\ohm}$ in our experiment). This gain circuit is powered by a $\pm 5~\si{\volt}$ power supply. For this configuration, the equivalent negative resistance parameter $R_\text{eq} \approx 2.58 \times 10^{-12}~\Omega$ using LTspice simulation. The nonlinear saturation behavior is simulated using two TVS diodes DF2S5M5SL and a resistor ($R_\text{non} = 510~\si{\ohm}$), which also limit the output voltage of the gain element.

\vspace{2em}
\backmatter

\bmhead{Data availability}
All data needed to evaluate the conclusions are presented in the article and/or the Supplementary Information. Raw data and corresponding simulation data are available upon reasonable request.

\bmhead{Code availability}
The relevant code is available from the first and corresponding authors upon reasonable request.

\bmhead{Supplementary information}
Supplementary information is provided.

\bmhead{Acknowledgements}
X.H. is grateful to Prof. Todor Mishonov for useful discussions. The work is supported by the National Natural Science Foundation of China (51977165), the Key Research and Development Program of Shaanxi Province (2024GX-YBXM-236), the Postdoctoral Fellowship Program of CPSF (GZB20240469), Hong Kong RGC Theme-Based Research Scheme Project (T23-701/20R) and Hong Kong RGC GRF (11205222). 

\bmhead{Author contributions}
X.H. and T.D. conceived the research. X.H., K.Y. and S.C. designed the experimental methodology. S.C., X.H. and R.W. conducted the experiment. X.H., K.Y. and T.D. performed the analysis. J.Z., X.M., C.K.Tse and T.D. supervised the project. X.H., K.Y. and T.D. wrote the original draft, and all authors discussed results and contributed to the writing, review, and editing of the article.

\bmhead{Declarations}
The authors declare that they have no competing interests.

\setlength{\bibsep}{0.5em plus 0.5ex}
\bibliography{main}

\vspace{2em}
\begin{flushright}
Dated: July 22, 2024
\end{flushright}
\end{document}